# Cooperative Negotiation in Autonomic Systems using Incremental Utility Elicitation

**Craig Boutilier**
Department of Computer Science
University of Toronto
Toronto, ON, M5S 3H5, Canada
cebly@cs.toronto.edu

**Rajarshi Das    Jeffrey O. Kephart**
**Gerald Tesauro    William E. Walsh**
IBM T.J. Watson Research Center
19 Skyline Dr.
Hawthorne, NY 10532, USA
rajarshi, kephart, gtesauro, wwalsh1 @us.ibm.com

## Abstract

Decentralized resource allocation is a key problem for large-scale autonomic (or self-managing) computing systems. Motivated by a data center scenario, we explore efficient techniques for resolving resource conflicts via cooperative negotiation. Rather than computing in advance the functional dependence of each element's utility upon the amount of resource it receives, which could be prohibitively expensive, each element's utility is elicited incrementally. Such incremental utility elicitation strategies require the evaluation of only a small set of sampled utility function points, yet they find near-optimal allocations with respect to a minimax regret criterion. We describe preliminary computational experiments that illustrate the benefit of our approach.

## 1 Introduction

The long-term goal of autonomic computing is to develop systems that can manage themselves with little or no human intervention [7]. Such systems must possess the ability to configure themselves, monitor performance and adapt to changing circumstances, self-optimize, and diagnose and repair problems. In large, distributed computing systems, such autonomy will generally require the continuous allocation and re-allocation of resources (e.g., compute cycles or storage) to distinct computing elements. As we elaborate below, the reasoning required to support optimal resource allocation is necessarily distributed, thus requiring some form of cooperative negotiation among the computing elements that have conflicting needs for critical resources.

To motivate our approach, we consider the task of an automated resource manager, or *provisioner*, allocating resources to various *workload managers (WMs)*. Each WM, given a specific allocation of resources, must decide how best to use those resources to service various client contracts. As a result, the utility of a specific allocation level to a WM often depends on the solution of a complex optimization problem. The provisioner's task is to allocate resources to the WMs in a way that total (organizational) utility is maximized. However, since the individual WM utility functions are complex and have no closed form—generally, even the computation of a single utility point in the WM function is complex and very expensive—it is infeasible to communicate entire utility functions directly to the provisioner.

We develop a model for distributed, cooperative negotiation in which the provisioner interacts with WMs through a form of incremental *utility elicitation*. In our model, the provisioner asks WMs for samples of their utility function at certain critical allocation levels. We describe techniques by which the provisioner can allocate resources based on this partial utility information. Because distributional information over utility functions is hard to obtain, we use a distribution-free model, *maximum regret,* to measure the quality of such an allocation. We describe computational methods for computing max regret, as well as methods for computing (and approximating) allocations with minimal max (minimax) regret. We also describe several elicitation methods that are guaranteed to offer improvement in decision quality in the worst case, and that, in practice, improve worst-case decision quality very quickly.

The remainder of the paper is organized as follows. In Section 2 we describe the resource allocation problem for a data center with multiple WMs, using this to motivate the more general model that follows. We argue that this problem should viewed as a form of cooperative negotiation, and solved using incremental utility elicitation, in Section 3. In Section 4, we formalize our model, and present exact and approximate algorithms to compute allocations with minimax regret given a set of partially known WM utility functions. In Section 5, we describe incremental elicitation strategies designed to reduce minimax regret, and present results demonstrating the effectiveness of these strategies in Section 6 using the data center model. We conclude in Section 7 with a discussion of future research directions.

## 2 Resource Allocation in an Autonomic System

We begin by describing the class of tasks that motivates this research, namely, the problem of resource allocation in autonomic systems. In this section, we provide a description of the basic task, while in Section 3 we argue that using incremental utility elicitation provides an appropriate means to facilitate the negotiation for resources among cooperative elements in an autonomic system. The formal details of our model will be introduced in Section 4.

An autonomic computing system is designed to drastically reduce the role of human administrators by automating most of the managerial decision making required in the operation of a complex computing environment [7]. Automated resource allocation, in particular, is necessary for an autonomic system to optimize its performance and adapt to failures that reduce resource availability. In large, distributed autonomic systems, resource allocation occurs at multiple scopes. Local allocation decisions will be made within individual elements (servers, databases, storage units, etc.) and small clusters of elements. Local clusters will contend for pools of resources in the larger domain, or across administrative domains. Although elements in an autonomic system of a single corporation will generally be cooperative (sharing the goal of optimizing total business value), the complexity of local information often precludes centralized allocation across the entire system. Cooperative negotiation, using preference elicitation techniques, can serve as an effective approach to decentralizing the problem.

To motivate the problem, consider resource allocation within a data center.[1] The center provides information technology resources to multiple organizations, separating domains for different groups of clients. Within a domain, resources are managed by a *workload manager*, such as IBM's enterprise Workload Manager (eWLM) [8]. Each WM decides how to allocate resources in its domain to maintain *quality of service (QoS)* for each of its transaction classes. The QoS specification for a transaction class is specified by a contract with customers, indicating monetary payments or penalties as a function of the QoS provided to transactions in the class. While a real WM may require multiple resource types, we assume for simplicity, in this paper, that the WM uses only a single, scalar type of resource. Given a distribution of its client demand, a model that maps the demand and the resource level to QoS, and the contract (which maps QoS to revenue), WM $i$ can compute $u_i(a_i)$, the maximum expected revenue it could obtain with the allocation of resource level $a_i$.

Because the distribution of client demand changes over time, the data center provisioner will periodically reallocate resources between the WMs. Letting $i$ range over WMs, the resource allocation problem for the provisioner is to compute (where $A$ is the set of feasible allocations, e.g., vectors of the form $\langle a_1, \ldots, a_n \rangle$):

$$\arg\max_{\mathbf{a} \in A} \sum_i u_i(a_i) \qquad (1)$$

---

[1] Our algorithms do not depend on the specific data center scenario we study. Indeed, the algorithms are applicable to a broad class of cooperative distributed allocation problems.

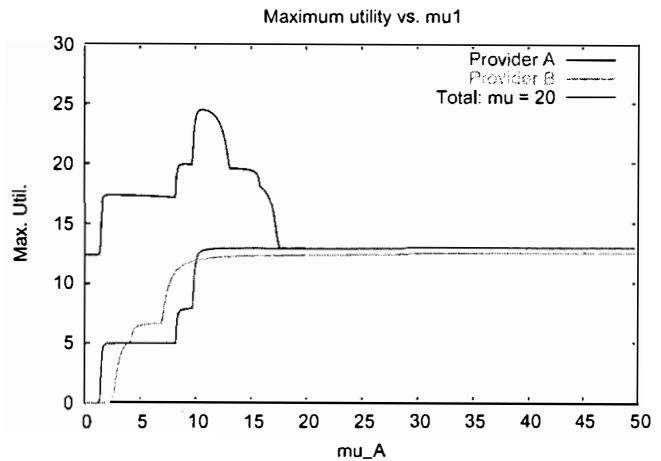

Figure 1: Maximum system utility as a function of allocation, with total resources $a = 20$. Curve "Provider A" indicates maximum utility to A as a function of $a_A$. Curve "Provider B" indicates maximum utility to B as a function of $a_B$. Curve "Total" indicates total utility as a function of $a_A$ provided to A (with $a - a_A$ provided to B).

The provisioner can compute Eq. 1 centrally if it has a good model of the internal operation of each WM and can obtain all relevant state information, including client demand distributions. In a real system, however, the model and data tend to be large and complex (this is certainly true in eWLM). Moreover, in a system with transient, heterogeneous components (e.g., differently configured WMs), the internal models of the components may simply be unavailable to the provisioner.

A very natural way to approach the problem of decentralized resource allocation is to have each WM determine its utility curve $u_i$, and communicate this function to the provisioner. With these functions in hand, the provisioner can determine an optimal allocation using a suitable optimization technique to solve Eq. 1. Since most of the information required to compute $u_i$ (e.g., queue arrival rates, dynamic QoS guarantees and pricing, etc.) is not directly available to the provisioner, communication of $u_i$ curves offers the most expeditious way of decomposing both computation and communication of relevant information between the WMs and the provisioner. Figure 1 shows an example of the utility curves of two WMs, each with two transaction classes.[2] The provisioner wishes to find the maximum of the aggregate (total utility) curve, also shown.

## 3 Modeling Cooperative Negotiation as Utility Elicitation

If a WM's contracts have a simple form and we have a simple QoS model (e.g., M/M/1 queue), then $u_i$ may have a tractable closed form (e.g., piecewise linear or quadratic). However, in typical systems the dependency of service attributes on resources and demand is sufficiently complex as to require a combination of optimization and simulation to compute $u_i$ at a *single* allocation point. Moreover, WMs will often have substitutable and complementary preferences over of *multiple* goods, giving rise to large, expensive-to-compute, multidimensional $u_i$ curves. Such complexities would make it infeasible for a WM to even compute its full $u_i$ curve, let alone communicate this to the provisioner.

Instead, we propose to model the resource allocation problem as cooperative negotiation. In the context of autonomic computing, cooperative negotiation is not simply non-cooperative negotiation with the simplifying assumption that agents are non-strategic. Rather, the objective is to achieve the right balance between global optimization effort, local computational expense, negotiation time, and decision quality. To this end, we treat the communication between the provisioner and the WMs as a form of *utility elicitation*. Specifically, our model allows the provisioner to ask each WM for its utility value for a small set of *sampled* allocations chosen by the provisioner to contain the most useful information with respect to determining an approximately optimal global allocation. In this sense, our work can be viewed in the same spirit as work on incremental utility elicitation [2, 4, 6, 10], where the aim is to obtain utility information that is most useful in improving decision quality. However, our model is very distinct.

Partial elicitation will generally be sufficient in negotiation among WMs. Given only a small number of samples of the utility functions $u_i(a_i)$, by making simple monotonicity assumptions, the provisioner can often determine the region of allocation space in which the optimal allocation lies.[3] For instance, having samples of the two (lower) $u_i$ curves in Figure 1 at points $a_i = \{10, 15, 20\}$ for $i = A, B$, is sufficient to determine that the optimal allocation lies some-where in the region $a_A \in [10, 15]$.

Knowing the region in which an optimal allocation lies is not enough. Given partial information in the form of sampled utility points, the provisioner must still decide on a specific allocation. Generally, no allocation can be guaranteed optimal since, for any allocation, there exists some utility function consistent with the sampled points for which that a better allocation exists. For this reason, we use the *minimax regret* decision criterion to compute allocations under utility function uncertainty [5]. This model bounds the error associated with the provisioner's allocation assuming an adversary picks a utility function, consistent with the current sampled points, in order to make the allocation as unattractive as possible. We develop this model in Section 4, and describe algorithms for computing (and approximately computing) allocations with minimax regret.

Minimax regret is a commonly used decision criterion in situations characterized by *strict uncertainty* [1, 5], that is, when uncertainty cannot be quantified probabilistically. Since distributional information over utility functions is hard to assess in the applications that currently motivate our model, we focus on the minimax regret criterion for optimization. If priors are available in a specific scenario, Bayesian techniques for optimization with imprecise utility information and utility elicitation [4, 2] could be used as well. We defer such a treatment to future work.

Given a specific set of sampled utility points from each WM, the regret associated with the minimax-optimal allocation may be too high. In this case, the provisioner has the opportunity to ask the WMs for additional sampled utility points. In Section 5, we describe elicitation strategies whose aim is to reduce minimax regret as quickly as possible. We describe several strategies, including a theoretically motivated method that provides worst-case guarantees on regret improvement, and heuristic methods that are more promising from a practical perspective (i.e., tend to get good results with far fewer queries). The reduction of minimax regret through incremental utility elicitation has been addressed previously [3, 9], though it does not appear to have been tackled in the context of complex cooperative negotiation.

The elicitation process is incremental: the provisioner obtains partial utility information from the WMs; it uses that information to determine a minimax-optimal allocation; and if this allocation has unacceptable error, the current utility samples are used to direct further queries. This process can be viewed as a form of cooperative negotiation, overseen by the provisioner, in which each WM reveals relevant information about its demands and expected revenue. Unlike typical economic mechanisms, the amount of revelation is limited, and focused on those areas of allocation space that are relevant to determining an optimal allocation.

---

[2] The QoS metric is response time. We computed the utility curves assuming a simple M/M/1 queue model.

[3] Monotonicity of $u_i$ is a natural assumption in this domain, corresponding to a "free disposal" assumption. One might be tempted to posit that, in certain scenarios, having additional resources can lead to lower expected utility (e.g., Braess's paradox comes to mind). However, assuming that a WM is simply interested in optimizing its own utility by the optimal use (or lack thereof) of allocated resources, the monotonicity assumption seems more than reasonable.

## 4 Minimax Regret

In this section we make the model more precise, and discuss the problem of allocating resources with partial utility information (specifically, sampled utility curves).

### 4.1 Sampled Utility Curves

We assume a provisioner charged with the task of allocating resources to a collection of $n$ WMs. To keep the presentation simple, we assume that the provisioner has some fixed amount of a single resource type to allocate. An *allocation* is a vector $\mathbf{a} = \langle a_1, \ldots, a_n \rangle$ such that $a_i \geq 0$ and $\sum_i a_i \leq 1$. Here $a_i$ refers to the *fraction* of the resources obtained by WM $i$. We denote by $A$ the set of feasible allocations. WM $i$'s *utility function* $u_i : [0, 1] \to \mathbf{R}$ associates a utility with any allocation of resources to it; specifically, $u_i(a_i)$ denotes the expected utility that WM $i$ will realize if it is given fraction $a_i$ of the resources under the control of the provisioner.[4] A *utility vector* $\mathbf{u} = \langle u_1, \ldots, u_n \rangle$ is a collection of such utility functions, one per WM.

We define the *value* of an allocation $\mathbf{a}$ under utility vector $\mathbf{u}$ to be the sum of the WM utilities:

$$V(\mathbf{a}, \mathbf{u}) = \sum_{i \leq n} u_i(a_i)$$

Notice that we make an implicit commensurability assumption, allowing the addition of WM utilities. In cooperative settings such as ours, this is generally acceptable (since, say, individual WM utility might measure its contribution to organizational value). We use the sum of individual utilities to reduce notational clutter; but arbitrary nondecreasing functions (transformations) can also be applied to the $u_i$, and these sums taken as well, without any substantive impact on our techniques.

Given a collection of utility functions $\mathbf{u}$, the provisioner is charged with the task of determining an *optimal allocation* w.r.t. $\mathbf{u}$: $\max_{\mathbf{a} \in A} V(\mathbf{a}, \mathbf{u})$. In general, this is a complex nonlinear optimization problem, since we make few assumptions about the structure of the individual $u_i$. A much more difficult problem emerges however: the construction of an optimal allocation requires full knowledge of the individual utility functions. But as mentioned above, even the calculation of a *single* utility point $u_i(a_i)$ by WM $i$ can be extremely difficult. Since $u_i$ will generally have no simple closed form, assuming full access to $u_i$ is problematic.

This difficulty can be circumvented if the provisioner is permitted to construct an approximately optimal allocation based on partial utility information. We assume: (1) Each WM can evaluate its utility function at specific points $a_i$, but cannot provide a closed form representation of $u_i$. (2) $u_i$ is monotonic non-decreasing (i.e., if $a_i \geq a'_i$, then $u_i(a_i) \geq u_i(a'_i)$), and that each WM can easily determine an upper bound on the fraction of resources $a_i^\top$ it can profitably use (i.e., it can find a point $a_i^\top$ s.t. $u_i(a_i) = u_i(a_i^\top)$ for all $a_i \geq a_i^\top$). (3) The provisioner can query each WM by providing an allocation level $a_i$ and receiving in response $u_i(a_i)$, the evaluation of $u_i$ at the query point.[5]

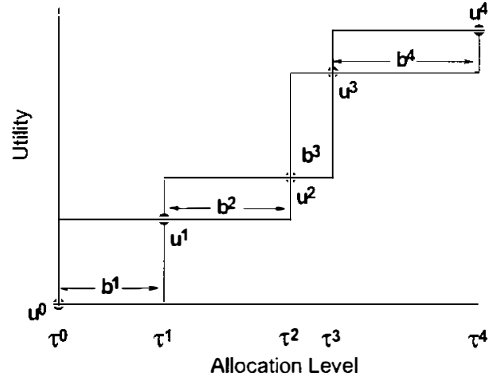

Figure 2: Bounds on the set of feasible utility functions.

### 4.2 Minimax Regret

We defer the question of elicitation to the next section. For now, we assume the provisioner has a collection of samples of each WM's utility function. Specifically, let

$$0 = \tau_i^0 < \tau_i^1 < \ldots < \tau_i^k = a_i^\top$$

be a collection of $k + 1$ thresholds at which samples $u_i(\tau_i^j)$ have been provided. We assume that the extreme utility values $u_i(0)$ and $u_i(a_i^\top)$ have been determined.[6] Notice that this collection of samples defines a set of $k$ *bins* into which an allocation of resources to WM $i$ can be placed. Allocation $a_i$ is said to lie within bin $b_i^j$ if $\tau_i^{j-1} < a_i < \tau_i^j$, in which case it has a lower bound on its utility of $u_i(\tau_i^{j-1})$ and an upper bound of $u_i(\tau_i^j)$. We use the notation $[a_i]$ to denote the index $j$ of the bin in which $a_i$ lies (if $a_i = \tau_i^j$ lies at a threshold, we let $[a_i] = j$). Hence, $b_i^{[a_i]}$ is the bin in which $a_i$ lies.

A utility function $u_i$ for WM $i$ is *feasible* (w.r.t. to the sample evaluations) iff it is nondecreasing and is consistent with the sampled points. A utility vector $\mathbf{u}$ is feasible iff each component $u_i$ is feasible. We denote by $U$ the set of feasible utility vectors given a set of utility samples $S$ ($S$ will generally be clear from context), and by $U_i$ the set of feasible utility functions $u_i$ for WM $i$. Figure 2 shows bounds on a WM utility function given a set of samples. The vertical lines indicate bin boundaries, and the horizontal lines upper and lower bounds on utility.

---

[4]WM $i$ will not have utility for such "fractions" explicitly, but rather for the amount corresponding to this share.

[5]We assume w.l.o.g. that $\sum_i a_i^\top \geq 1$. If this is false, the provisioner has more resources that the WMs can use jointly and the optimization problem faced by the provisioner is trivial.

[6]For notational convenience, we assume that $k+1$, the number of sampled points, is the same for all WMs $i$—this is simply to keep subscripting to a minimum (nothing depends on this).

Given incomplete knowledge of WM utility functions in the form of samples, the provisioner can measure the quality of a specific allocation in terms of its *maximum regret*. This gives a bound on the worst-case error associated with an allocation, assuming an adversary can pick the true utility vector from the feasible set $U$.

**Definition** The *maximum regret of allocation* **a** *w.r.t. allocation* **a'** is

$$MR(\mathbf{a}, \mathbf{a}') = \max_{\mathbf{u} \in U} V(\mathbf{a}', \mathbf{u}) - V(\mathbf{a}, \mathbf{u})$$

The max regret of allocation **a** is then

$$MR(\mathbf{a}) = \max_{\mathbf{a}' \in A} MR(\mathbf{a}, \mathbf{a}')$$

An allocation $\mathbf{a}^* \in \arg\min_{\mathbf{a} \in A} MR(\mathbf{a})$ is said to have *minimax regret*. The minimax regret level $MMR(U)$ of feasible utility set $U$ is $MR(\mathbf{a}^*)$.

Minimax regret offers a reasonable method for resource allocation in the face of utility function uncertainty. It minimizes the amount of utility one could sacrifice by acting in the face of such uncertainty. We refer to an allocation with minimax regret as *minimax optimal*.

### 4.3 Computing Max Regret

There is a single feasible utility function $u_i$ (for each WM) that gives $MR(\mathbf{a}, \mathbf{a}')$ (w.r.t. any competing allocation $\mathbf{a}'$). We set the utility over the interval $[\tau_i^{[a_i]-1}, a_i]$ to the lower bound $u_i(\tau_i^{[a_i]-1})$, and the interval $(a_i, \tau_i^{[a_i]}]$ to the upper bound. All other bins $b_i^j$ are set to their maximum values. (If $a_i = \tau_i^j$ for some $j$, then all utilities are set to their upper bounds). The utility vector **u** obtained by applying this to each $u_i$ gives **a** its least possible feasible utility. It also gives every other allocation maximum utility, with the exception of those allocations $a'_i \leq a_i$ that lie within the same bin as $a_i$. But for any such allocation, we have $u_i(a'_i) - u_i(a_i) \leq 0$ by monotonicity, and this vector ensures this quantity is 0. Thus, **u** maximizes the regret of **a**. w.r.t. any other allocation $\mathbf{a}'$.

Determining max regret $MR(\mathbf{a})$ thus requires searching for an allocation $\mathbf{a}^w$ that maximizes $V(\mathbf{a}^w, \mathbf{u}) - V(\mathbf{a}, \mathbf{u})$. We call this allocation a *witness* for **a**. This witness can be computed using a mixed integer program (since **u** is no longer a "variable"). Specifically, assume two sets of variables: $A_i$ ($i \leq n$) is a real-valued variable denoting the allocation to WM $i$; and $B_i^j$ ($i \leq n, j \leq k$) is a 0-1 variable denoting that the allocation to $i$ lies in bin $b_i^j$. We then solve the MIP:

*Maximize:* $\quad \sum_{i \leq n, j \leq k} B_i^j u_i^j$

*subj. to:* $\quad 0 \leq A_i \leq a_i^\top; \quad \sum_i A_i \leq 1$

$\quad\quad\quad 1 \leq \forall j \leq k, \quad \sum_i B_i^j = 1,$

$\quad\quad\quad \forall j > 1, \quad A_i/(\tau_i^{j-1}) - B_i^j \geq 0 \quad\quad (2)$

$\quad\quad\quad \forall j < k, \quad B_i^j - \frac{a_i^\top - A_i}{a_i^\top - \tau_i^j} \leq 0 \quad\quad (3)$

Constraints (2) and (3) constrain the lower bound of $a_i$'s bin to be $\tau_i^{[a_i]-1}$ and the upper bound of $a_i$'s bin to be $\tau_i^{[a_i]}$, respectively. The $\tau_i^j$ thresholds are the usual with the exception that $\tau_i^{[a_i]-1}$ (i.e., the lower bound of $b_i^{[a_i]}$) is replaced with $a_i$. The utilities $u_i^j$ are defined to be the upper or lower bounds associated with these bins, as defined in the construction above.

Once a regret-maximizing witness $\mathbf{a}^w$ has been determined, the max regret of **a** is just the difference in utility of $\mathbf{a}^w$ and **a** given worst-case utility function for **a**.

### 4.4 Computing Minimax Regret

Several exact and heuristic strategies can be used to find an allocation with minimax regret. We describe several such techniques here. Key to our methods is the notion of a pointwise allocation. Assume a set of sampled points $S_i = \{\tau_i^j\}$ for each WM $i$. A *pointwise allocation* **p** is any allocation such that $p_i \in S_i$, for all $i \leq n$. In other words, WM $i$ is given a fraction of resources at which it has provided a utility sample. An *exhaustive pointwise allocation (EPA)* is any feasible pointwise allocation that cannot be extended by allocating more resources to any WM in a way that feasibly attains a new pointwise allocation. More precisely, **p** is an EPA if it is pointwise, $\sum_i p_i \leq 1$, and for all $i'$, $\tau_{i'}^{[p_{i'}]+1} + \sum_{i \neq i'} p_i > 1$.

The *supporting pointwise allocation* for **a**, $SPA(\mathbf{a})$, is the pointwise allocation $\mathbf{a}^p$ whose threshold values are those at or just below the allocation values of **a**. In other words, if $\tau_i^j \leq a_i < \tau_i^{j+1}$, then $\mathbf{a}_i^p = \tau_i^j$. For any EPA **p**, denote by $E(\mathbf{p})$ the set of *extensions* of **p**, that is, the set of exhaustive allocations whose SPA is **p**. Pointwise allocations have a fixed utility (i.e., there is no uncertainty about their utility). It is not hard to see that $MR(\mathbf{a}) \leq MR(SPA(\mathbf{a}))$. In addition, any allocation **a** can be written as $SPA(\mathbf{a}) + \boldsymbol{\delta}$, where $\delta_i \geq 0$. We call $\delta(\mathbf{a}) = \sum_i \delta_i$ the *surplus* associated with **a**. The worst case utility of **a** is equal to the (fixed) utility of $SPA(\mathbf{a})$.

We can restrict our attention to exhaustive allocations (i.e., where $\sum_i a_i = 1$) in the search for minimax optimal allocations—this is a simple consequence of monotonicity. Furthermore, it isn't hard to see that minimax optimal allocations must lie among the set of exhaustive allocations whose SPA is an EPA. We will show that

$$MMR(E(\mathbf{p})) \equiv \min_{\mathbf{a} \in E(\mathbf{p})} MR(\mathbf{a})$$

can be computed effectively in an iterative fashion, for any EPA $\mathbf{p}$. As a result, since EPAs are enumerable, we can search through this set to find the allocation that achieves $MMR(E(\mathbf{p}))$ for each EPA $\mathbf{p}$, and be assured that the minimax optimal allocation is that with minimum max regret among this finite set of allocations. We now describe the computation of $MMR(E(\mathbf{p}))$.

Let $\mathbf{p}$ be an EPA. First assume that $\sum \mathbf{p}_i = 1 - \delta$ for some $\delta > 0$. If not, then $E(\mathbf{p}) = \{\mathbf{p}\}$ and $MMR(E(\mathbf{p})) = MR(\mathbf{p})$. We call $\delta$ the *surplus* of $\mathbf{p}$. Any exhaustive allocation in $E(\mathbf{p})$ has the form $\mathbf{a} = \mathbf{p} + \boldsymbol{\delta}$, where $\delta_i \geq 0$ is the portion of the surplus allocated to WM $i$ by $\mathbf{a}$, and $\sum_i \delta_i = \delta$.

Let $\mathbf{a}$ be any allocation in $E(\mathbf{p})$, and let $\mathbf{a}^w$ be a witness (that is, an allocation that maximizes regret for $\mathbf{a}$, obtained by solving the MIP described above). If $\mathbf{a}^w$ has the property that $a_i^w$ is not in the interval $(p_i, a_i]$ for any $i$, then $MR(\mathbf{a}) = MR(\mathbf{p})$. Intuitively, this holds because $\mathbf{a}^w$ must also be a witness for $\mathbf{p}$. Thus by computing minimax regret for $\mathbf{a}$ using the MIP described above, we obtain an upper bound on $MMR(E(\mathbf{p}))$; and if the solution of this MIP provides an allocation that has the property above, we are assured that $MMR(E(\mathbf{p})) = MR(\mathbf{p})$.

If this property does not hold, we can tighten this upper bound on $MMR(E(\mathbf{p}))$ as follows. Again, let $\mathbf{a}^w$ be the witness for $\mathbf{a}$, and let $\gamma = \delta(\mathbf{a}^w)$ be its surplus. Let $\mathbf{a}' = \mathbf{p} + \boldsymbol{\delta}'$ be any allocation in $E(\mathbf{p})$. The maximum pairwise regret $MR(\mathbf{a}', \mathbf{a}^w)$ is exactly

$$MR(\mathbf{p}, \mathbf{a}^w) - \sum \{u_i(\tau_i^{[a_i^w]+1}) - u_i(\tau_i^{[a_i^w]}) : b_i^{[a_i^w]} = b_i^{[a_i']}, \gamma_i \leq \delta_i'\}$$

Thus, the regret of any $\mathbf{a}'$ w.r.t. to the witness $\mathbf{a}^w$ for $\mathbf{a}$ is equal to the max regret of $\mathbf{p}$ w.r.t. $\mathbf{a}^w$ less the regret contributed by allocations of $\mathbf{a}^w$ to those $i$ where the new allocation $\mathbf{a}'$ exceeds that of $\mathbf{a}^w$, but both lie in the same bin.

Note that the regret of $\mathbf{a}'$ is maximized by *any* allocation $\mathbf{a}^w$ that allots all of its surplus (less some infinitesimal amount) to the those WMs $i$ where both $a_i^w$ and $a_i'$ lie in the same bin. This means that if $\gamma$ (the surplus of $\mathbf{a}^w$) is (strictly) greater than $\delta$ (the surplus of $\mathbf{a}'$), no matter how $\delta$ is distributed among the bins of $\mathbf{a}'$, we can allocate the surplus $\gamma$ of $\mathbf{a}^w$ among the same bins so that $\gamma_i > \delta_i$. This ensures that the max regret of $\mathbf{a}'$ is exactly that of $\mathbf{a}$. If $\delta \geq \gamma$, this is not possible. Specifically, let $m$ be the largest integer such that $\delta \geq m\gamma$. Then the allocation $\mathbf{a}' \in E(\mathbf{a}^p)$ that minimizes regret w.r.t. some $\mathbf{a}^w$ of the form above is obtained by allocating $\gamma$ (of the total $\delta$) of the surplus to the $m$ bins of $\mathbf{a}'$ that have the largest utility gaps (i.e., difference between their upper and lower bounds). This allocation ensures that $a_i^w$ cannot exceed $a_i'$ in any of these bins.

Hence the surplus $\gamma$ associated with $\mathbf{a}'$ must be allocated to the remaining $n - m$ lowest gapped bins. Thus we get a (generally tighter) upper bound on $MMR(E(\mathbf{a}^p))$:

$$MR(\mathbf{p}, \mathbf{a}^w) - \sum \{u_i(\tau_i^{[a_i]+1}) - u_i(\tau_i^{[a_i]}) : b^i \text{ is among}$$
$$\text{the } m \text{ bins with largest such gaps in utility}\}$$

This process finds the $\mathbf{a}' \in E(SPA(\mathbf{a}))$ that has minimal max regret w.r.t. *any* $\mathbf{a}^{w'} \in E(SPA(\mathbf{a}^w))$. Note that $\mathbf{a}^{w'}$ may not be a true witness for $\mathbf{a}'$—it simply maximizes the regret of $\mathbf{a}'$ among those allocations that have the same SPA as the witness $\mathbf{a}^w$ for $\mathbf{a}$. However, this process can be repeated with the new allocation $\mathbf{a}'$: we find its witness, and re-allocate its surplus in the same fashion.

This process is guaranteed to converge on an allocation that realizes $MMR(E(\mathbf{p}))$. The process terminates when either: the witness at the current iteration has the same SPA as a prior witness; the surplus associated with the current witness exceeds the surplus of current allocation; or the witness and the allocation share no bins. In each such case, the current allocation is minimax optimal in $E(\mathbf{p})$. The process only continues if $\mathbf{a}$ and $\mathbf{a}'$ share at least one bin, and the surplus for $\mathbf{a}'$ is less than that of $\mathbf{a}$. It can be shown that the max regret of the new allocation is no greater than that of the prior allocation. Since this procedure can consider only a finite number of distinct witness points (at most one per EPA), it is guaranteed to converge.

Note that upper and lower bounds on $MMR(E(\mathbf{a}^p))$ for each $\mathbf{a}^p$ can be produced rather easily. Thus any intelligent search scheme can be used to search through the space of pointwise allocations without enumerating them explicitly (e.g., using a branch-and-bound procedure). Furthermore, with such upper and lower bounds, one can determine an approximately minimax-optimal allocation as well.

We note that several other heuristic methods for generating minimax allocations can be used. For example, an optimal *optimistic* allocation $\mathbf{a}^o$ can be computed using an IP: one simply sets the WM utility functions to their upper bounds and finds the best allocation. Let $\varepsilon_{\max}$ be the largest *utility gap* associated with sample set $S$:

$$\varepsilon_{\max} = \max_i \max_j u_i(\tau_i^{j+1}) - u_i(\tau_i^j)$$

Then $MR(\mathbf{a}^o) \leq 2\varepsilon_{\max}$ for any optimistic $\mathbf{a}^o$.

Rather than relying on an IP, greedy search methods can be used as well. One simple technique generates an EPA incrementally by increasing allocation of exactly one WM to its next threshold value. Initially, we let each WM be assigned zero resources. A move can be made in this search space by increasing one of the WM's allocation from its current level $\tau_i^j$ to $\tau_i^{j+1}$ as long as the total allocation remains feasible. A simple heuristic for evaluating moves is $\frac{u_i(\tau_i^{j+1}) - u_i(\tau_i^j)}{\tau_i^{j+1} - \tau_i^j}$ (i.e., increase in marginal utility per unit of resource). Again, since *true* max regret of any allocation can be computed readily, even when we approximate, we have guaranteed regret bounds.

## 5 Elicitation Strategies

We turn our attention to the question of elicitation. We assume the provisioner has a collection $S$ of sampled utility points from the WMs, and has computed a minimax optimal allocation $\mathbf{a}(S)$ (or some approximation thereof). If the provisioner is unhappy with the regret $MR(\mathbf{a}(S), S)$, it can ask utility queries of any of the WMs to obtained additional sampled utility points.

We describe two strategies, one theoretically motivated to perform well in the worst-case (i.e., when an adversary chooses WM utility functions, hence responses to our queries), and one based on more practical intuitions that we expect to work well in practice.

We start with an analysis of worst-case behavior. For simplicity, assume that $a_i^\top = 1$ for each WM $i$ (i.e., each WM can profitably use all available resources). Let $u_i(1) = \varepsilon_i$. After a single query to each WM, we have a single bin for each WM with lower bound 0 and upper bounds $\varepsilon_i$, hence a *utility gap* of size $\varepsilon_i$. Each query of WM $i$ can be seen as dividing this original bin into smaller bins with smaller utility gaps. However, an adversary can choose a utility function that ensures regret never goes to zero.

**Proposition 1** *There exists a set of WM utility functions $u_i$ such that with no finite number of queries can minimax regret be reduced below $\frac{\max \varepsilon_i}{2}$.*

Intuitively, such a set requires that each WM utility function be arbitrarily close to a step function, that jumps from utility level 0 to $\varepsilon_i$ at some critical resource level. With certain restrictions on either the discrete allocation of resources, or the first derivative of the individual $u_i$, this problem can be circumvented.

Despite the fact that regret may never reach zero, this lower bound on minimax regret can be achieved in a polynomial number of queries using a simple "halving" procedure. The halving procedure asks a sequence of queries of each WM such that each bin is divided in half. Specifically, let $k = 2^m - 1$ for some $m$; after $k$ queries, the utility samples for a specific $u_i$ will consist of $k+1$ bins of size $\frac{1}{k+1}$. After $O(n^2)$ such queries of each WM, we are assured that regret can be no more than the lower bound above.

**Proposition 2** *The halving procedure, after no more than $2n(n-1)$ queries of each WM, results in a collection of utility function samples whose minimax regret is no more than $\frac{\max \varepsilon_i}{2}$ (where $\varepsilon_i$ is the original utility gap of WM $i$).*

It is important to emphasize that an adversary must pick a very specific utility function for each WM to ensure this worst-case bound. In practice, the halving strategy focuses effort on parts of utility space that are not relevant to determining minimax optimal allocations.[7]

An intuitively simple strategy that works much better in practice is based on the intuition that to reduce regret, we want to improve the information we have about the allocation currently estimated to be optimal. Specifically, we'd either like to show that its regret is less than currently estimated (in which case we've improved our ability to make an good decision, by reducing minimax regret), or gain information that will help us make a better decision. Given $\mathbf{a}$ and $\mathbf{a}^w$, the provisioner asks each WM $i$ for its utility value at points in $p_i \in b_i^{[a_i]}$ or $p_i^w \in b_i^{[a_i^w]}$. If $p_i \leq a_i$ and the response $u_i(p_i)$ is greater than its lower bound, the max regret of $\mathbf{a}$ must be reduced; similarly, if $p_i^w \geq a_i^w$ and the response $u_i(p_i^w)$ is less than its upper bound. In either case, minimax regret will be reduced. If each WM responds with its upper bound in the case of $u_i(p_i^w)$ and its lower bound for $u_i(p_i)$, then the max regret of the current best allocation has not, unfortunately, been reduced. However, we note that a great deal of uncertainty in the each of the WM's utility functions has been removed. Specifically, if a WM ever responds to a query with the upper (resp., lower) bound on its value, then all uncertainty is removed from the range of utility values between the query point and the upper (resp., lower) threshold of the bin in which it lies.

Because it is expensive to query both $p_i$ and $p_i^w$, in our experiments we chose to query either $p_i$ or $p_i^w$ according to the following heuristic. The heuristic value of a bin $b_i^j$ is the sum of the scaled height and width of the bin, defined as $\Delta u + \Delta \tau$, where $\Delta u \equiv (u_i(\tau_i^j) - u_i(\tau_i^{j-1}))/u_i(a_i^\top)$ and $\Delta \tau \equiv (\tau_i^j - \tau_i^{j-1})/a_i^\top$. We queried the bin (allocation or witness) with the highest heuristic value. We also tried querying just the allocation bin and just the witness bin. We found that the former did not reduce regret quickly and the latter reduced regret comparably to the heuristic, except when the number of query points grew larger, in which case the heuristic performed better.

Although an obvious choice for $p_i^w$ is exactly $a_i^w$, we found it did not work well in practice because $a_i^w$ was often very close to the low end of the bin $\tau_i^{[a_i^w]-1}$. We found that uncertainty is reduced more quickly when we chose $p_i^w$ at the midpoint of the bin (and similarly for $p_i$).

To reduce computation time in our experiments, we computed approximately optimal allocations. For each exhaustive pointwise allocation we compute the max regret of a small number (1-3) of random extensions, and choose the extension with the minimum max regret over all EPAs. The basic procedure we used in our experiments then consists of the following steps: (a) query each WM $i$ for its utility at $a_i = 0$ and $a_i = 1$, as well as two more randomly chosen points; (b) determine an approximate minimax optimal allocation a as well as its witness $\mathbf{a}^w$ using the current utility samples, terminating if $MR(\mathbf{a})$ is below some acceptable threshold; (c) otherwise, ask one query of each WM $i$ at the midpoint of bin $b_i^{[a_i^w]}$ of bin $b_i^{[a_i]}$; and repeat with the increased sample set.

---

[7]Indeed, we believe that a conditional variant of the halving strategy, where only specific bins are halved depending on the current samples, can attain our worst-case bound with a logarithmic number of queries. We do not yet have a proof of this however.

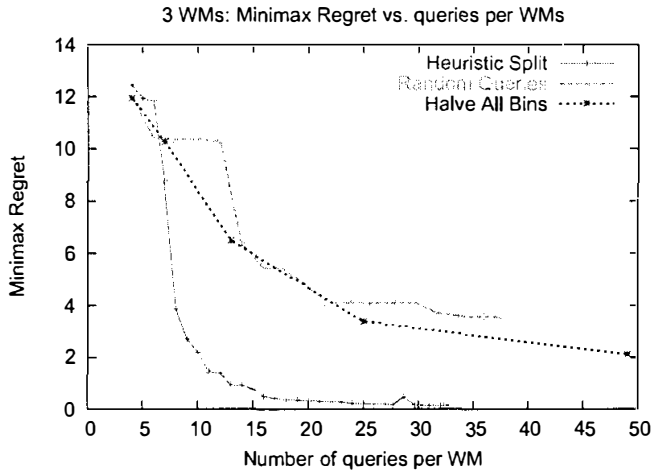

Figure 3: Minimax regret for 3 WMs as a function of number of queries per WM for our strategy "Heuristic Split" and two alternative query strategies, "Random Queries" and "Halve All Bins."

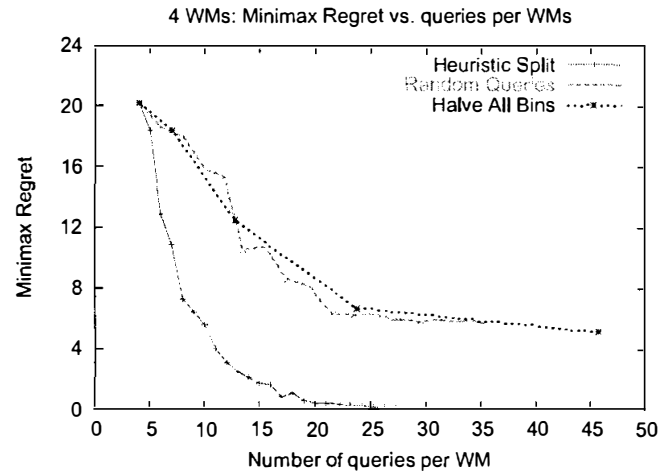

Figure 4: Minimax regret for 4 WMs as a function of number of queries per WM for our strategy "Heuristic Split" and two alternative query strategies, "Random Queries" and "Halve All Bins".

## 6 Empirical Results

This section describes results of our elicitation strategy for our data center model. We studied configurations with three and four WMs, each with two transaction classes. Client contracts specified payments as a function of response time. The functions were (roughly) slightly smoothed out step functions, with high payments for response time below a threshold, and zero payments above the threshold. Given a fixed level of resource, a WM controls the response time of each class through the fraction of available resource assigned to that class. We used a simple M/M/1 queue to model response time in each class.

Our numerical implementation of the MIP computation of max regret utilizes the GNU Linear Programming Kit (GLPK) version 3.2.4. While GLPK solves for a regret-maximizing witness as a continuous variable, we constrain all queries to lie on a discretized grid of 10000 points in the unit interval. This discretization makes it easier to compute an individual WM's maximal utility for a given resource level, and also eliminates floating-point roundoff errors in identifying the bin containing a given allocation.

Our implementation also uses a bounding procedure to greatly reduce the number of MIPs computed. During the loop over the EPAs, we keep track of the current best witness seen so far. We can quickly compute the regret of any other allocation with respect to the current best witness, giving us a lower bound on the max regret for the allocation. If this lower bound is greater than the lowest max regret found so far, we know that the allocation cannot be the minimax regret allocation. We found that most EPAs can be eliminated as minimax-optimal on this basis, without actually invoking the MIP computation of max regret, resulting in a reduction in CPU time by nearly a factor of 100.

Figures 3 and 4 plot sample runs illustrating typical behavior of our elicitation strategy (denoted "Heuristic Split") for three and four WMs, respectively. For comparison purposes, we also plot the (approximate) minimax regret values obtained using two less intelligent querying strategies: "Random Queries" generates random queries drawn from a uniform distribution in the unit interval, while "Halve All Bins" is the halving procedure discussed previously in Proposition 2. In both figures, the minimax regret of our strategy decreases rapidly with the number of queries, demonstrating the effectiveness of our approach. Furthermore, our strategy achieves significantly lower minimax regret values than the other strategies, for a given number of query points per WM.

Plots of the data in Figures 3 and 4, using a log scale for the vertical axis, show a reasonably linear decrease for our algorithm. This suggests that our procedure is able to reduce minimax regret exponentially with the number of queries. While the alternative algorithms will generally reach zero regret with a sufficient number of queries, the rate of de-crease is much slower than exponential.

The minimax regret of our strategy reduced more slowly with four WMs than with three. This is not surprising, since the space of joint utility functions, and hence the total uncertainty, grows with the number of WMs.

Our tentative results for two, three and four WMs suggest a scaling of GLPK CPU time of roughly $N^q$, where $N$ is the number of WMs and $q$ is the number of queries per WM. As a consequence, we found it to be computationally prohibitive to compute the minimax regret for larger ($q > 15$) number of queries with more than four WMs. With respect to computation time, we must emphasize, however, that our goal is to minimize the amount of utility information that each WM must provide, since determining a single utility point requires intensive computation on the part of a WM. In addition, these values will generally change over time, requiring re-elicitation and re-allocation (which is one of the main motivations for the autonomic model). Thus we generally see the number of queries per WM being rather small. Furthermore, preliminary tests suggest we can obtain more than an order of magnitude speedup with state-of-the-art MIP solvers such as CPLEX. Finally, we expect that the more computationally feasible heuristic strategies suggested in Section 4.4 will prove to be extremely valuable as a means of generating queries.

## 7 Concluding Remarks

We have argued that cooperative negotiation using incremental utility elicitation is required to perform resource allocation in a distributed autonomic system. To address this need, we presented algorithms for computing minimax regret, and two elicitation strategies: a blind bin halving strategy and a strategy that halves the bins of the minimax-optimal allocation and its regret-maximizing witness. We empirically demonstrated, in a data center provisioning scenario, that the more directed strategy quickly reduces minimax regret. Furthermore, although we could demonstrate a theoretical guarantee of max regret convergence for the blind strategy, the heuristic strategy performs much better in practice.

In future work we will develop faster (and possibly more approximate) minimax regret algorithms to enable the study of larger problems. We suggested a greedy strategy that may not compute the minimax optimum allocation but saves computation because it requires that max regret be computed only once. However, it is not acceptable to approximate the max regret computation, as then we would lose any known guarantee on the quality of the allocation. Ultimately though in the context of autonomic computing our focus is on the cost of *elicitation*, hence we need to better understand the tradeoff between acceptable levels of minimax regret and the cost of elicitation.

We intend to expand our model to include multidimensional utility for multiple resources. The concomitant increase in utility space will generally result in greater utility uncertainty. Further algorithm developments will likely be necessary to achieve acceptable regret levels without an explosion in the requisite number of preference queries.

We also plan to study elicitation strategies for Bayesian optimization criteria. Bayesian approaches may reduce the number of needed queries by providing value of information guidance as well as tighter bounds on the value of an allocation (i.e., expected value of an allocation, rather than worst-case bounds). To use Bayesian techniques, a provisioner must form a prior distribution over the WMs' utility functions, which may require the provisioner to employ learning techniques along with models of the internal operation of WMs, as well as WM and client demand dynamics. Such models must be fairly minimal though because of the constraints imposed by decentralization.

**Acknowledgments:** Craig Boutilier gratefully acknowledges the support of an IBM Faculty Partnership Award.